\documentclass{emulateapj}

\slugcomment{}
\usepackage{graphicx}
\usepackage{longtable}

\shorttitle{A Very Large  Galaxy at $Z=3.72$}
\shortauthors{Lee et al. }
\begin{document}
\def\hh{\, h^{-1}}
\newcommand{\ie}{$i.e.$,}
\newcommand{\wth}{$w(\theta)$} 
\newcommand{\mpc}{Mpc}
\newcommand{\xir}{$\xi(r)$}
\newcommand{\Lya}{Ly$\alpha$}
\newcommand{\Lyb}{Lyman~$\beta$}
\newcommand{\Hb}{H$\beta$}
\newcommand{\msun}{M$_{\odot}$}
\newcommand{\hmsun}{$h^{-1}$M$_{\odot}$}
\newcommand{\sfr}{M$_{\odot}$ \text{yr}$^{-1}$}
\newcommand{\dnsty}{$h^{-3}$Mpc$^3$}
\newcommand{\za}{$z_{\rm abs}$}
\newcommand{\ze}{$z_{\rm em}$}
\newcommand{\cmtwo}{cm$^{-2}$}
\newcommand{\nhi}{$N$(H$^0$)}
\newcommand{\degpoint}{\mbox{$^\circ\mskip-7.0mu.\,$}}
\newcommand{\halpha}{\mbox{H$\alpha$}}
\newcommand{\hbeta}{\mbox{H$\beta$}}
\newcommand{\hgamma}{\mbox{H$\gamma$}}
\newcommand{\kms}{\,km~s$^{-1}$}      
\newcommand{\minpoint}{\mbox{$'\mskip-4.7mu.\mskip0.8mu$}}
\newcommand{\mv}{\mbox{$m_{_V}$}}
\newcommand{\Mv}{\mbox{$M_{_V}$}}
\newcommand{\peryr}{\mbox{$\>\rm yr^{-1}$}}
\newcommand{\secpoint}{\mbox{$''\mskip-7.6mu.\,$}}
\newcommand{\sqdeg}{\mbox{${\rm deg}^2$}}
\newcommand{\squig}{\sim\!\!}
\newcommand{\subsun}{\mbox{$_{\twelvesy\odot}$}}
\newcommand{\et}{{\it et al.}~}
\newcommand{\er}[2]{$_{-#1}^{+#2}$}
\def\h50{\, h_{50}^{-1}}
\def\hbl{km~s$^{-1}$~Mpc$^{-1}$}
\def\ltsima{$\; \buildrel < \over \sim \;$}
\def\simlt{\lower.5ex\hbox{\ltsima}}
\def\gtsima{$\; \buildrel > \over \sim \;$}
\def\simgt{\lower.5ex\hbox{\gtsima}} 
\def\arcs{$''~$}
\def\arcm{$'~$}
\newcommand{\wu}{$U$}
\newcommand{\wb}{$B_{435}$}
\newcommand{\wv}{$V_{606}$}
\newcommand{\wi}{$i_{775}$}
\newcommand{\wz}{$z_{850}$}
\newcommand{\hmpc}{$h^{-1}$Mpc}
\newcommand{\lm}{$L$--$M$}
\newcommand{\ws}{$\mathcal{S}$}
\newcommand{\wm}{$\mathcal{M}$}
\newcommand{\sm}{$\mathcal{S}$-$\mathcal{M}$}
\newcommand{\medianLM}{$\tilde{\mathcal{L}}(M)$}
\newcommand{\mf}{$\phi_\mathcal{M}$}

\title{Discovery of a Very Large  ($\approx 20$~kpc) Galaxy at $z=3.72$ }

\author{Kyoung-Soo Lee\altaffilmark{1,12},  Arjun Dey\altaffilmark{2}, Thomas Matheson\altaffilmark{2}, Ke Shi\altaffilmark{1}, Chao-Ling Hung\altaffilmark{3}, Rui Xue\altaffilmark{4},  Hanae Inami\altaffilmark{5}, \\Yun Huang\altaffilmark{1}, Khee-Gan Lee\altaffilmark{6}, Matthew L.~N. Ashby\altaffilmark{7}, Buell Jannuzi\altaffilmark{8}, Naveen Reddy\altaffilmark{9}, Sungryong Hong\altaffilmark{10}, \\ Wenli Mo\altaffilmark{11},  Nicola Malavasi\altaffilmark{1}}
\altaffiltext{1}{Department of Physics and Astronomy, Purdue University, 525 Northwestern Avenue, West Lafayette, IN 47907}
\altaffiltext{2}{National Optical Astronomy Observatory, Tucson, AZ 85726}
\altaffiltext{3}{Department of Physics, Manhattan College, 4513 Manhattan College Parkway, Riverdale, NY 10471 }
\altaffiltext{4}{Department of Physics and Astronomy, University of Iowa, Iowa City, IA 52242}
\altaffiltext{5}{Observatoire de Lyon, 9 avenue Charles Andre, Saint-Genis Laval Cedex F-69561, France}
\altaffiltext{6}{Lawrence Berkeley National Laboratory, 1 Cyclotron Rd., Berkeley, CA 94720}
\altaffiltext{7}{Harvard-Smithsonian Center for Astrophysics, 60 Garden St., Cambridge, MA 02138}
\altaffiltext{8}{Steward Observatory, University of Arizona, Tucson, AZ 85721}
\altaffiltext{9}{Department of Physics and Astronomy, University of California, Riverside, 900 University Avenue, Riverside, CA 92521, USA}
\altaffiltext{10}{School of Physics, Korea Institute for Advanced Study, 85 Hoegiro, Dongdaemun-gu, Seoul 02455, Republic of Korea}
\altaffiltext{11}{Department of Astronomy, University of Florida, Gainesville, FL 32611}
\altaffiltext{12}{Visiting astronomer, Kitt Peak National Observatory (KPNO), National Optical Astronomy Observatory, which is operated by the Association of Universities for Research in Astronomy (AURA) under a cooperative agreement with the National Science Foundation.}
\altaffiltext{}{This paper is based on observations obtained at the Gemini Observatory, which is operated by the Association of Universities for Research in Astronomy, Inc., under a cooperative agreement with the NSF on behalf of the Gemini partnership: the National Science Foundation (United States), the National Research Council (Canada), CONICYT (Chile), Ministerio de Ciencia, Tecnolog\'{i}a e Innovaci\'{o}n Productiva (Argentina), and Minist\'{e}rio da Ci\^{e}ncia, Tecnologia e Inova\c{c}\~{a}o (Brazil) (Program GN-2017A-DD-2).
}
\begin{abstract}
We report the discovery and spectroscopic confirmation of a very large star-forming Lyman Break galaxy, G6025, at $z_{\rm spec}=3.721\pm0.003$.  In the rest-frame $\approx$2100\AA, G6025 subtends $\approx$24~kpc in physical extent when measured from the 1.5$\sigma$ isophote, in agreement with the parametric size measurements which yield the half-light radius of $4.9\pm0.5$~kpc and the semi-major axis of $12.5\pm0.1$~kpc.  G6025 is also very UV-luminous ($\approx 5L^*_{\rm UV,z\sim4}$) and young ($\approx 140\pm60$~Myr). Despite its unusual size and luminosity, the stellar population parameters and dust reddening ($M_{\rm{star}}\sim M^*_{z\sim4}$, and E($B-V$)$\sim$$0.18\pm0.05$) estimated from the integrated light, are similar to those of smaller galaxies at comparable redshifts. The ground-based morphology and spectroscopy show two dominant components, both located off-center, embedded in more diffuse emission. We speculate that G6025 may be a scaled-up version of chain galaxies seen in deep HST imaging, or alternatively,   a nearly equal-mass merger involving two super-$L^*$ galaxies in its early stage. G6025 lies close to but not within a known massive protocluster at $z=3.78$. We find four companions within  $ 6$~Mpc from G6025, two of which lie within 1.6~Mpc. While the limited sensitivity of the existing spectroscopy does not allow us to robustly characterize the local environment of G6025, it likely resides in a locally overdense environment.  The luminosity, size, and youth of G6025 make it uniquely suited to study the early formation of  massive galaxies in the universe.
\end{abstract}
  

\section{Introduction}

In the canonical picture of galaxy formation, the growth of galaxies is closely linked to that of  dark matter halos \citep{wr78}. Recent observations  generally support  this expectation.  Galaxy sizes increase with cosmic time at rates similar to those of halos \citep{ferguson04,bouwens04,law12,shibuya15,ribeiro16}. The positive correlation between the  size, stellar mass, and star formation rate of star-forming galaxies hint at the profound impact that halo assembly histories have on galaxy formation at all cosmic epochs \citep{shen03,franx08,mosleh11, mosleh12, law12, targett13, ribeiro16}.

The current theoretical expectation is that both external and internal processes play central roles in shaping the key properties of galaxies. The hierarchical picture of galaxy growth invariably requires frequent mergers and interactions, particularly at high redshift. Moreover, young gas-rich star-forming disks are dynamically unstable, leading to  fragmentation into giant clumps characterized by turbulent kinematics \citep[e.g.,][]{bournaud07,dekel09}.
Indeed, the structural properties of high-redshift star-forming galaxies are characterized by clumpy morphologies \citep{ravindranath06,lotz06,law12, wuyts12, guo12, guo15, guo17}, high ellipticities \citep{elmegreen05}, and high velocity dispersion \citep{law07,forsterschreiber09,wisnioski15}, markedly different from those of low-redshift disk galaxies. Multiplicities (i.e., galaxies consisting of multiple `knots') appear to be a common feature operating in all luminosity and mass ranges \citep[e.g.,][]{elmegreen05clump,elmegreen07,hodge13,targett13}.

Despite the current ambiguity in the relative importance of external and internal processes in the evolution of these galaxies, it is clear that massive star-forming galaxies are ultimately headed to form systems resembling local disk galaxies. At $z\sim2-3$, the highest star-formers are best described by an exponential surface brightness profile  \citep{law12,targett13}. The fraction of galaxies with disky morphologies increases while measured velocity dispersion decreases with cosmic time \citep{forsterschreiber09,wisnioski15}; these trends signal that the stellar disk is being transformed into a stable and rotationally supported one. Together, these observations paint a fairly coherent picture of disk formation in the cosmological context. 

In this paper, we report the discovery of a very large star-forming galaxy at $z=3.721$, which we name  G6025. In many ways, G6025 appears to defy the general expectation of galaxy formation as described above. The morphology of G6025 -- based on ground-based data -- resembles a disk galaxy viewed edge-on, but it is too large in size  given the large look-back time at which it is observed. Moreover, despite its unusually large size, the stellar population parameters lie in the range of a typical galaxy at high redshift, making it a significant outlier in the size-mass scaling law. While some of the radio galaxies and submillimeter-bright galaxies at high redshift are known to have similarly large sizes \citep[e.g.,][]{vanbreugel98,targett13}, they are 1-2 orders of magnitude brighter than G6025, hinting at a different  origin of their formation. 
 Finally, its spatial proximity to one of the most massive protoclusters  discovered to date \citep{lee14,dey16} is also curious.

The unusual properties of G6025 likely reflect that it represents  an extremely short-lived phase of galaxy formation, or it may signify an entirely different pathway to form massive galaxies operating specifically in dense environments.   While  we are unable to conclusively determine its physical nature without better data, its exceptional brightness in the rest-frame UV wavelengths will enable future observations to obtain a useful insight into one of the main modes of galaxy growth in exquisite detail. Here, we present the current observational constraints on this unique object.

We use the WMAP7 cosmology $(\Omega, \Omega_\Lambda, \sigma_8, h) = (0.27, 0.73, 0.8, 0.7)$ from \citet{wmap7}. Distance scales are given in physical units unless noted otherwise. All magnitudes are given in the AB system \citep{oke83}. In the adopted cosmology, 1\arcsec\ corresponds to the angular diameter distance 7.4~kpc at $z=3.721$.

\section{Photometric Data}

\begin{figure*}
\epsscale{1.1}
\plotone{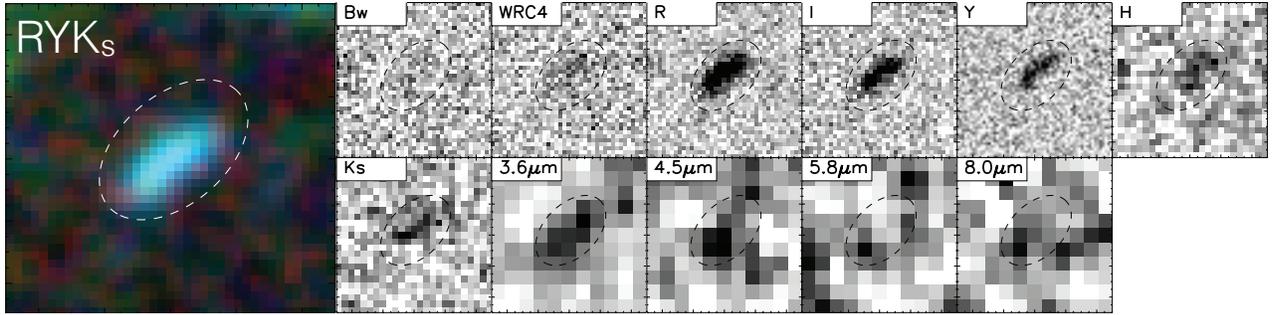}
\caption{
{\it Left:} a false-color $RYK_S$ image centered on G6025. {\it Right:} postage stamp images of G6025 (from left) in the $B_W$, $R$, $I$, $Y$, $H$, $K_S$, IRAC $3.6\mu m$,  $4.5\mu m$, $5.8\mu m$, and $8.0\mu m$ bands. All images are 10\arcsec\ on a side. The single-band  images are displayed in their native pixel scales, while the color image is created from the resampled, PSF-matched data. 
Image orientation is such that north is up and east is to the left. In each panel, an ellipse of constant angular size ([$a$,$b$]=[2\farcs7,1\farcs7]) is displayed as a visual aid. }
\label{stamps}
\end{figure*}

G6025 was discovered during our ongoing investigation of a massive protocluster at $z=3.78$ named PC217.96+32.3 \citep{lee13,lee14}. The structure is  currently one of the largest and most overdense regions known in the high redshift universe. Based on the observed galaxy overdensity, we estimate that it will collapse into systems of masses exceeding $>10^{15}M_\odot$ by the present day \citep{dey16}, making it a Coma cluster analog observed billions of years prior to its final coalescence. 

A complete description of multi-wavelength data acquisition in the PC217.96+32.3 field and reduction procedures will be presented in an upcoming paper (K. Shi et al., in prep), and here we only provide a brief description. 
The optical data were taken with four broad-band filters ($B_WRIY$: NOAO program IDs: 2012A-0454, 2014A-0164) using the Mosaic camera on the Mayall telescope \citep{jacoby98,mosaic3} and the Hyper SuprimeCam on the Subaru telescope \citep{hsc}. In the same period, narrow-band imaging data were obtained using the WRC4 filter (designed to sample C~{\sc iv} emission in Wolf-Rayet stars; KPNO filter no.~k1024) which has a central wavelength of 5819\AA\ and a FWHM of 42\AA\ \citep{lee14}. The details of the Mosaic observations and characteristics of the final mosaics are given in \citet{lee14}.  The near-infrared $HK_S$ data were obtained using the NEWFIRM instrument \citep{newfirm1,newfirm2} on the Mayall telescope (NOAO Program IDs: 2015A-0168, 2016A-0185).  The existing {\it Spitzer} IRAC data from the Spitzer Deep Wide-Field Survey \citep{ashby09} were reprojected to the same astrometric grid as the rest of the dataset.  

All optical and near-IR images were resampled to a common pixel scale of 0\farcs258\,pix$^{-1}$, the native pixel scale for the $B_WRI$ and $WRC4$ bands.   The native pixel scale of the $Y$ and $HK_s$ bands, respectively 0\farcs168 and 0\farcs4, made this feasible for those bands.  The native pixel scale for the {\it Spitzer}/IRAC arrays is however much larger, 1\farcs2, so we chose to resample our IRAC mosaics to 0\farcs777\,pix$^{-1}$, three times coarser than the rest of the data.  It was necessary to resample the IRAC data to a pixel scale that is an integer multiple of that used for the other optical data in order to reliably extract optimal photometry with our template-fitting method (see below).


The $5\sigma$ limiting magnitudes measured in a 2\arcsec\ diameter aperture are 26.88, 26.19, 25.37, and 25.10~AB in the optical data ($B_WRIY$), 24.05 and 24.83~AB in the near-IR data ($HK_S$), respectively. The seeing measured in the stacked images is 1\farcs0 in the $B_WRI$ images, 0\farcs6 in the $Y$-band, and 1\farcs2 in the near-IR bands. 

For each band, we use the PSFEx software \citep{psfex} to measure the image point spread function (PSF) out to 3\arcsec\ from peak brightness. Taking the worst seeing data as the target PSF, a two-dimensional  convolution kernel is derived for each image, and is used for PSF homogenization. In deriving the kernels, we use the full shape of the observed stellar profiles rather than assuming a functional form such as Moffat profiles. The details of the PSF matching procedure  are given in \citet{xue17}. 

Source detection and photometric measurements were carried out using the SExtractor software \citep{bertina96} on the PSF-matched images where the $K_S$-band data was used as detection image. For the IRAC data, PSF-matched photometry was performed using the TPHOT software \citep{tphot1,tphot2}; the unconvolved $K_S$ band image and its source list were used for the flux fitting. Finally, these catalogs were combined into a single multi-wavelength catalog of our survey field. The catalog includes photometry for all $K_S$-band detected sources within a contiguous 28\arcmin$\times$28\arcmin\ field.  The SExtractor parameter MAG\_AUTO is used to estimate the total magnitude, while colors are computed from fluxes within a fixed isophotal area (i.e., FLUX\_ISO). Colors measured using FLUX\_ISO and FLUX\_APER (3\arcsec\ diameter circular) are in agreement with each other within 0.1~mag. Since the images are PSF-matched, aperture correction is given by the difference between MAG\_AUTO and MAG\_ISO estimated in the $K_S$ band.  

\section{Discovery and  Confirmation of G6025}

G6025 was initially identified as an unusually large galaxy. The strong spectral break  implied from the red $B_W-R$ and blue $R-I$ color ($B_W-R\gtrsim 3.2$, $R-I=0.18\pm0.14$) suggests that it lies at $z\sim 3.3-4.1$.  In Figure~\ref{stamps}, we show postage stamp images of G6025 in all photometric bands. Each image has a dimension of 10\arcsec\ on a side, and is displayed in its native seeing and pixel scale. 

 The apparent linear extent of G6025 exceeds 3\arcsec\ in five photometric bands (from $R$ to $K_S$) with its major axis running in the NW-SE direction with the position angle of  303.75 degree. A false-color image constructed from the PSF-matched $RYK_S$ data shows no discernible color gradient across its face (Figure~\ref{stamps}, left). The mean colors are $R-Y=0.22\pm0.10$, $Y-K_S=1.07\pm0.15$; within the photometric uncertainties, the colors of the upper and lower half are identical.  There may be a change occurring between the 3.6$\mu$m and 4.5$\mu$m bands where the upper half of G6025 dims. However, given the coarse image resolution and shallow depth of the IRAC bands, the change is marginally significant.
We determine a best-fit photometric redshift using the CIGALE software \citep{burgarella05,noll09,cigale} to be $z_{\rm{phot}}=3.80\pm0.27$. This is close to the spectroscopic redshift of the protocluster PC217.96+32.3 at $z=3.78$ \citep{dey16}.  

We search for G6025's counterpart in several existing datasets which cover the region. We check the 24$\mu$m, 70$\mu$m, and 160$\mu$m data of the MIPS AGN and Galaxy Evolution Survey \citep[MAGES:][]{jannuzi10} provided by \citet{vaccari16}. We also cross-match its position to the source lists compiled from several radio observations, including the 62~MHz LOFAR  \citep{vanweeren14}, VLA 324.5~MHz image \citep{coppejans15}, and VLA 1.4~GHz observations from the FIRST survey \citep{becker95}\footnote{ The 5$\sigma$ limiting sensitivities are 0.23, 14.86, and 80.30~mJy for the MIPS 24, 70, and 160~$\mu$m bands, and   24, $\sim1-4$, and 0.75~mJy beam$^{-1}$ for the radio data in the 62~MHz, 324.5~MHz, and 1.4~GHz, respectively.}.  No counterpart is identified in all of our searches. While non-detection suggests that G6025 is not a system dominated by the central blackhole activity, one cannot entirely rule out the possibility of AGN presence given the shallow depth of the data. 


\subsection{Optical spectroscopy of G6025}
We obtained spectroscopy of G6025 on the Gemini North telescope (Program ID: GN-2017A-DD-2). The observations were carried out on UT 2017 April 5 using the newly upgraded GMOS-N with the Hamamatsu CCDs. We used R400\_G5305 grating blazed at 764 nm. A long slit (1\arcsec\ width) was placed with the position angle of 303.75\,deg  aligned with the major axis of the source. The science observations consisted of 9 individual 1200~sec exposures, resulting in the total integration time of 3 hours. Individual frames were dithered in both spectral and spatial direction. For the spectral dithers, we used two wavelength settings (centered on 600~nm and 605~nm) to cover the gap between the CCDs and get full wavelength coverage. We used 15\arcsec\ offsets for spatial dithers. The standard star Feige 34 was observed for flux calibration using the identical setup.

We use IRAF\footnote{IRAF is distributed by the National Optical  Astronomy Observatory, which is operated by the Association of  Universities for Research in Astronomy, Inc., under cooperative agreement with the National Science Foundation.} and Gemini-IRAF for standard CCD processing and spectrum extraction. The spectrum covers $\sim4000-8400$\AA.  We use an optimized version\footnote{https://github.com/cmccully/lacosmicx} of the LA Cosmic algorithm \citep{vandokkum01} to eliminate cosmic rays. Sky is removed from each frame by fitting to a second-order polynomial. Based on these data, we create two different average stacks using {\tt iraf/imcombine}: (1) a combination of all sky-subtracted frames; and (2) a combination of the images created by differencing pairs taken with identical spectral but different spatial dithers. The latter image provides a cleaner sky subtraction at the expense of reduced signal-to-noise ratio. We extract a one-dimensional spectrum from each 2D spectrum.   We employ our own IDL routines to flux calibrate the data, and to remove telluric lines using the well-exposed continua of the spectrophotometric standards \citep{wade88,matheson00}.

\begin{figure*}
\epsscale{1.2}
\centering
\plotone{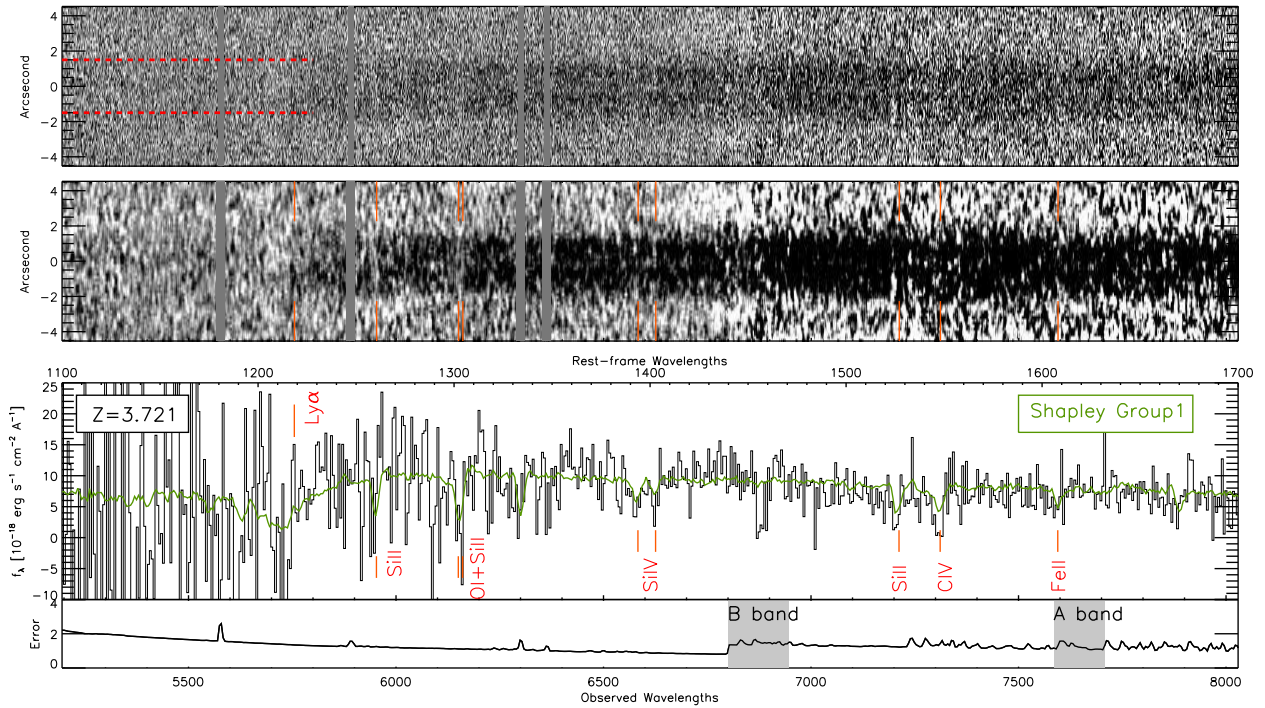}
\caption{
{\it Top:} the two-dimensional spectrum of G6025: the spatial direction is indicated in units of arcsecond.  Four bright sky lines are marked as grey bands.  The extraction aperture (3\arcsec) is indicated by red dashed lines on left.
{\it Middle:} the same spectrum is smoothed by a two-dimensional boxcar ($N=3\times 5$). It is apparent that several absorption features run across the entire extent of the galaxy, including Si~{\sc ii}$\lambda$1260, O~{\sc i}+Si~{\sc ii}$\lambda$1303, and Si~{\sc iv}$\lambda\lambda$1393,1402.
{\it Bottom:} the 1D spectrum and error spectrum are shown together with the \citet{shapley03} `Group 1' LBG spectrum redshifted to $z=3.721$ (green line).  Red lines show redshifted wavelengths of interstellar (IS) absorption features present in the spectrum. Tentative detection of Ly$\alpha$ emission is redshifted relative to the interstellar absorption lines.
}
\label{spec}
\end{figure*}

In the top panel of Figure~\ref{spec}, we show the composite 2D spectrum of G6025, created by joining the average-stack spectrum on the blue side ($\lambda \leq 6800$\AA) and the difference stack spectrum on the red side.  While the signal is  diffuse, it is clear that the continuum breaks at 5700-5800\AA. There is a possible hint of very weak Ly$\alpha$ emission in the SE part of the spectrum. The spatial extent of the continuum is measured to be $\approx 3$\arcsec, consistent with the imaging data. The 2D spectrum shows two ridges surrounded by a diffuse continuum at $\lambda\lesssim 7000$\AA.  In the 2D spectrum, the local minimum occurs at  $y\approx 0$\arcsec\ where the intensity decreases down to $\approx 20$\% of the peak intensity. Even though the spectrum appears to change to a more uniform surface brightness at longer wavelengths, it is possible that it is simply due to increased noise rather than a real effect. 

In order to identify spectral features more clearly, we smooth the spectrum by applying a two-dimensional boxcar kernel ($N$=3 and 5 in the spatial and spectral direction, respectively), as shown in the middle panel of Figure~\ref{spec}. The pixel scale is 0.16\arcsec/pixel and 1.5\AA/pixel in the spatial and spectral direction, respectively.  In the bottom panel, we show the  one-dimensional spectrum extracted within a full 3\arcsec\ aperture.

For G6025, we measure a redshift of  $z=3.721\pm0.003$ based on the interstellar absorption lines. Very weak Ly$\alpha$ emission is tentatively detected redshifted (by $\approx 700$~km/s) with respect to several detections of interstellar absorption lines, including  Si~{\sc ii}$\lambda$1260, O~{\sc i}+Si~{\sc ii}$\lambda$1303, Si~{\sc ii}$\lambda$1526, C~{\sc iv}$\lambda\lambda$1548,1550, and at a lower significance, Si~{\sc iv}$\lambda\lambda$1393,1402. In particular, the Si~{\sc ii}$\lambda$1260 and O~{\sc i}+Si~{\sc ii}$\lambda$1303 doublet are clearly present across the entire spatial direction, unambiguously confirming that the entire extent of the luminous source lies at a common redshift and is not the result of line-of-sight projection of unrelated galaxies.

\begin{figure*}[th!]
\epsscale{1.0}
\plottwo{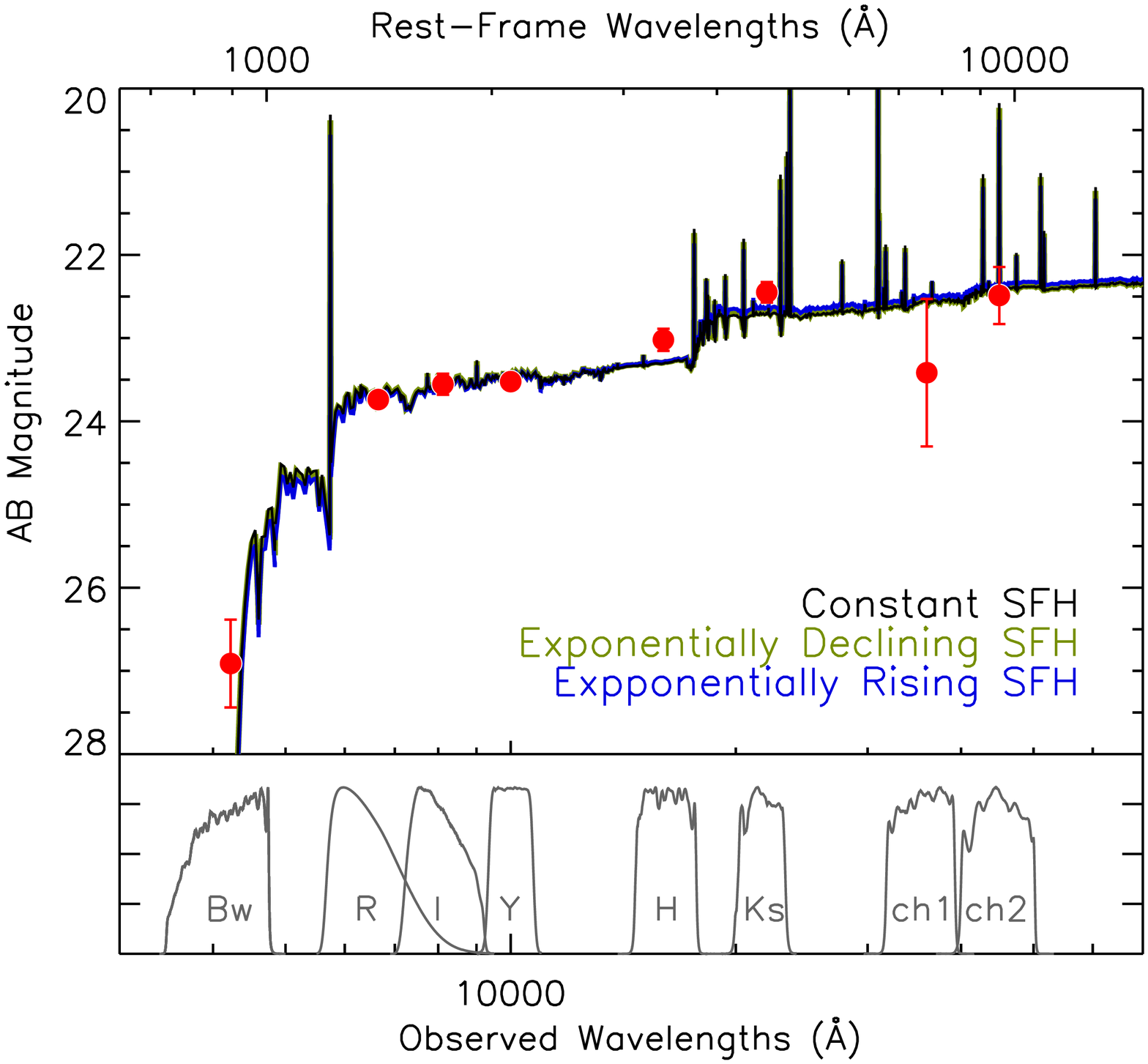}{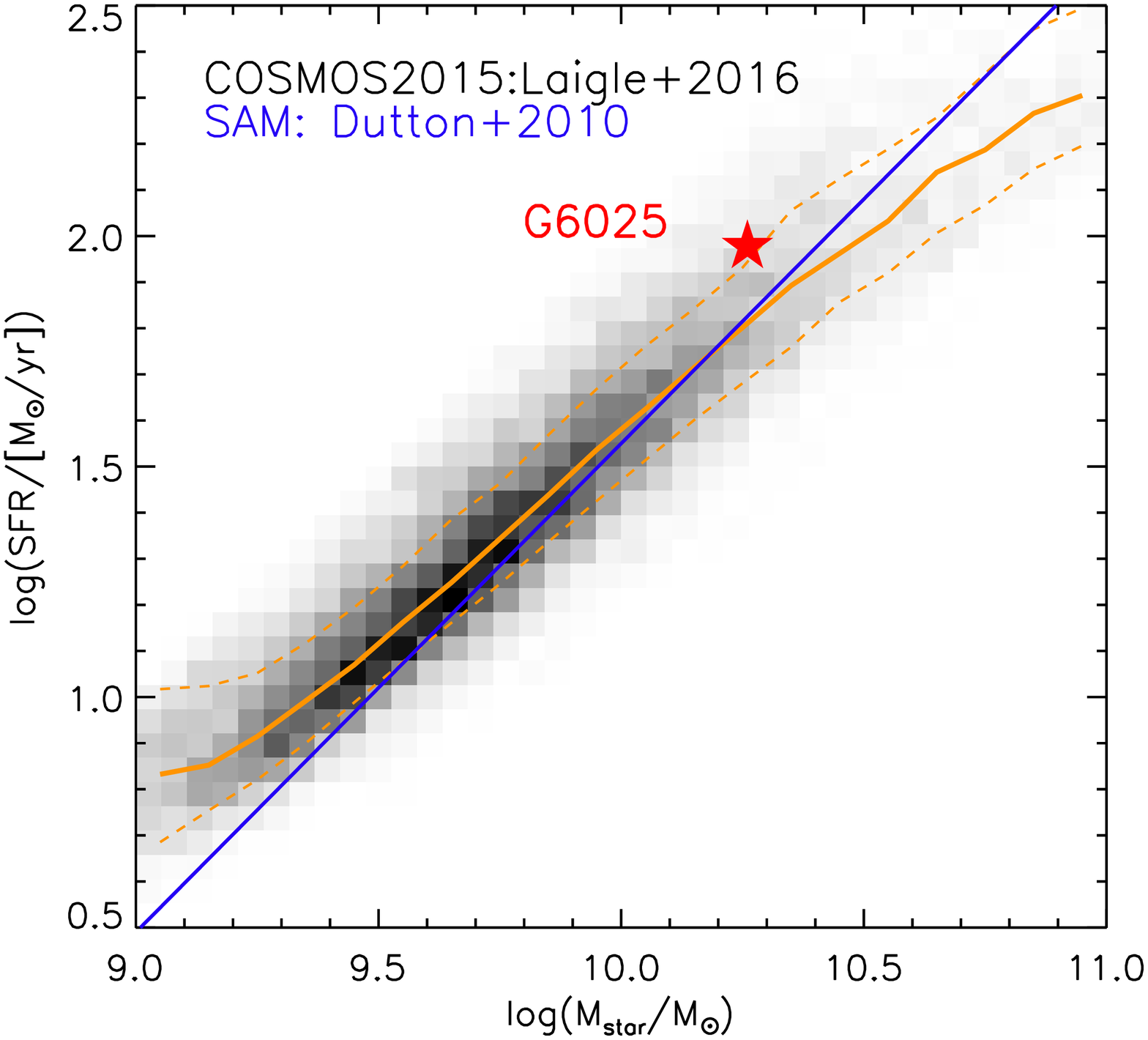}
\caption{
{\it Left:} Three best-fit SED models are shown together with the photometric measurements of G6025 (red data points). The response functions of the filters used in the fit are shown on bottom. 
{\it Right:} the $M_{\rm{star}}$-SFR scaling relation is for the star-forming galaxies at $z_{\rm{phot}}$=3.4-4.1 in the COSMOS field \citep{laigle16}. The source density in each bin is indicated by the  greyscale while the median SFR and its $1\sigma$ scatter (i.e., 16th and 84th percentile of the SFR distribution in a given stellar mass bin) are marked in orange solid and dashed lines, respectively. A red star marks the location of G6025. The predicted scaling relation from a semi-analytical model by \citet{dutton10} is shown in blue solid line.
}
\label{fig_sed}
\end{figure*}

While the overall shape of the G6025 spectrum is similar to  the composite spectrum of UV-luminous star-forming galaxies at $z\sim3$ \citep[][]{shapley03},  the interstellar absorption appears to be stronger in G6025 than typical $z\sim3$ galaxies. In Figure~\ref{fig_sed}, we overlay the Shapley `Group 1' composite spectrum for comparison (green line). The spectrum is constructed from a subset of $z\sim3$ galaxies with the lowest quartile of the  Ly$\alpha$ equivalent width distribution, which also has the strongest absorption features. For the absorption line equivalent widths for the O~{\sc i}+Si~{\sc ii}$\lambda$1303 doublet, Si~{\sc iv}$\lambda\lambda$1393,1402 doublet, and C~{\sc iv}$\lambda\lambda$1548,1550, we measure $-3.0\pm0.4$\AA, $-3.7\pm0.2$\AA, and $-5.5\pm0.5$\AA, respectively. For reference, the EWs found for the Shapley Group 1 sample are $-3.24\pm0.16$\AA, $-2.84\pm0.29$\AA, and $-3.56\pm0.30$\AA, respectively.  The average values for the \citet{shapley03} full sample are  $-2.20\pm0.12$\AA, $-2.64\pm0.23$\AA, and $-3.03\pm0.21$\AA. We find that  C~{\sc iv}$\lambda\lambda$1548,1550 absorption appears to be particularly strong ($\gtrsim 3\sigma$) compared to the Group 1 spectrum. With deeper spectroscopic observations scheduled in Spring 2018, we intend to investigate more detailed spectroscopic properties of G6025. 

\section{The Physical Properties of G6025}

\subsection{Stellar Population Parameters}
We derive the stellar population parameters, including stellar mass, dust reddening, and star formation rate (SFR) using the CIGALE software. We use the stellar population models of \citet{bc03}, the \citet{calzetti00} extinction law, solar metallicity, and \citet{chabrier03} initial mass function. Separate runs are made using three different star formation histories (SFHs): i) a constant star formation history (CSF); ii) an exponentially declining SFH; and iii)  an exponentially rising SFH. For the declining and rising SFHs, multiple $e$-folding timescales ($\tau$=[0.1,0.3, 0.5, 1.0] Gyr, where $\rm{SFR}\propto e^{\pm t/\tau}$)  are used to obtain a best-fit model. We allow redshift to float in one case, while fixing it to the measured redshift in others. The best-fit photometric redshift solution yields $z_{\rm phot}=3.80\pm0.27$, consistent with the spectroscopic redshift. At $z=3.721$, H$\beta$, [O~{\sc iii}], and H$\alpha$ lines redshift to 2.30$\mu$m, 2.36$\mu$m, and 3.10$\mu$m, respectively.  Only H$\beta$ falls into the red end of the $K_S$ filter, while the other two are well outside of the $K_S$ and 3.6$\mu$m band filter responses. Taking the median H$\beta$ equivalent width of $z$=3-4 galaxies given by \citet[][$W_0\leq 65$\AA]{schenker13}, the contamination in the  $K_S$ band is $<0.08$~mag: i.e., smaller than the photometric error itself. No correction was made to the photometry prior to the SED fitting.

\begin{table}
\caption{Key physical properties of G6025}
\begin{tabular}{lcccc}
\hline
\hline
 & CSF & CSF  & exp. decl. & exp. ris.\\
 &    (floating $z$)      & ($z$ fixed) & ($z$ fixed) & ($z$ fixed)\\
\hline
$z_{\rm{phot}}$ & $3.80\pm0.27 $& 3.72 & 3.72 & 3.72 \\
$\log{M_{\rm{star}}}$ &  $10.26^{+0.12}_{-0.16}$& $10.26^{+0.11}_{-0.14}$ & $10.31^{+0.11}_{-0.15}$ & $10.30^{+0.19}_{-0.13}$ \\
SFR & $97\pm31$ & $95\pm29$ & $51\pm31$ & $75\pm49$ \\
Age & $142\pm65$ & $144\pm64$ & $136\pm58$ & $144\pm74$ \\
E($B-V$) & $0.17\pm0.06$ & $0.18\pm0.05$ & $0.12\pm0.07$ & $0.14\pm0.07$ \\
$\chi_r^2$ &  4.69 & 4.83 & 4.74 & 4.81 \\
\hline
\end{tabular}
\tablecomments{SFRs and $M_{\rm{star}}$ are in units of $M_\odot$~yr$^{-1}$ and $M_\odot$, respectively. Stellar population age is in units of Myr.}
\label{stellar_pop}
\end{table}

The best-fit model parameters  are listed in Table~\ref{stellar_pop}.  Overall, young stellar ages are preferred as best-fit models tend to have large  $\tau$ values for the declining  SFH and small $\tau$ values for the rising SFH. All three SFHs considered give comparably good fits to the photometric data as shown in the left panel of Figure~\ref{fig_sed}. The stellar mass, dust reddening, and luminosity-weighted population age remain virtually unchanged regardless of the assumed SFHs.   The inferred SFR, on the other hand, varies up to 40\% reflecting the rapidly changing stellar-mass-to-UV-light ratio of the young stellar population.

The SED modeling suggests that G6025 is actively forming stars at the rate of 50-100~$M_\odot$~yr$^{-1}$, and contains dust similar to most $z\sim4$ galaxies   \citep{finkelstein12,bouwens14}. Its stellar mass is $\approx 2\times10^{10}M_\odot$, once again, typical of star-forming galaxies; the characteristic mass of the stellar mass function measured for star-forming galaxies is $\log{M^*_{\rm{star}}/M_\odot}=10.35^{+0.25}_{-0.30}$ at $z\approx 3.7$ \citep{gonzalez11,lee12a,grazian15,song16}. The bulk of these stars were formed  in the last $\sim140$ Myrs, placing the formation redshift at $z_f\approx 4$. Adopting the Salpeter IMF instead would increase the mass by 0.24~dex, while assuming a subsolar metallicity would lead to 10\% higher SFR, a slightly higher dust reddening ($\Delta$E($B$-$V$)=0.04), and younger age by $\sim$10~Myr. 

To assess how the physical properties of G6025 compare with the general population of star-forming galaxies,  we define a `control sample' using the COSMOS15 catalog \citep{laigle16}. We select the sources whose best-fit photo-z lie in the range  $z_{\rm{phot}}$=3.4-4.1. At $3<z<6$, the photometric redshift precision is estimated to be $\sigma_z/(1+z)=0.021$. After removing galaxies whose photo-z probability density functions are multiply peaked, 19,318 sources are selected. 
The CIGALE software was run on theses galaxies with the identical setup as previously (solar metallicity, Chabrier IMF, \citet{calzetti00} extinction law, and constant SFH). Redshift is fixed to the best-fit photometric redshift. Clearly, it is unrealistic to expect that all galaxies meet these assumptions. However, our simplistic assumptions are  justified by the fact that we are mainly interested in the relative position of G6025 on this parameter space rather than exploring the full behavior of galaxies in general. Because a SFH of a galaxy cannot be well constrained, a different choice of the adopted SFH would not change the location of G6025 on the SFR-$M_{\rm star}$ plane relative to the control sample.
\begin{figure*}
\epsscale{1.2}
\plotone{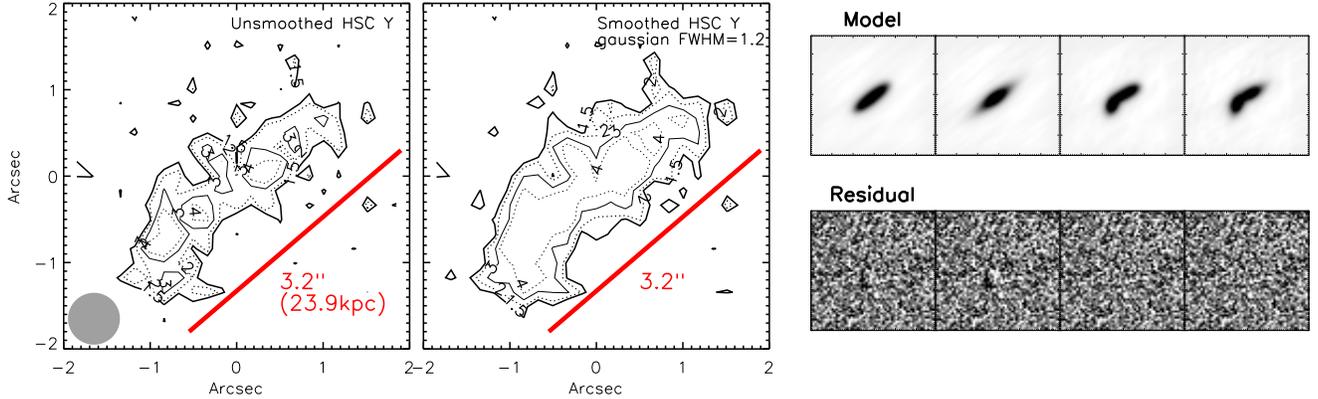}
\caption{
{\it Left:} The angular extent of G6025 is illustrated as pixel intensity contours at the 1.5$\sigma$, 2$\sigma$, 3$\sigma$ and 4$\sigma$ levels. The sky pixel-to-pixel rms noise, $\sigma$, is measured directly from the image after a number of sigma-clipping iterations. The contours on the unsmoothed image (left panel) show a complex structure with at least six $\geq 3\sigma$ peaks along the major axis (defined by thick red line). The angular size of a 1.5$\sigma$ isophotal area is 3.2\arcsec. Grey circle denotes the full-width-at-half-maximum  size of the seeing disk {\it Middle: } the $Y$-band image is smoothed using a gaussian kernel with the FWHM of 1.2 pixels; sky noise is recomputed after the smoothing. The size of a 1.5$\sigma$ isophote is comparable to that measured in the unsmoothed image, while the size of a 3$\sigma$ isophote is 2.9\arcsec. {\it Right:} Four best-fit GALFIT models and corresponding residual images are shown, from left to right, the single-component S\'{e}rsic, single-component exponential disk  ($n=1$),  two-component S\'{e}rsic , and two-component exponential disk models, respectively.
}
\label{fig_contour}
\end{figure*}

In the right panel of Figure~\ref{fig_sed}, we show their locations on the SFR-$M_{\rm{star}}$ plane. The source density in each bin is indicated by the greyscale while the median and $1\sigma$ scatter (i.e., 16th and 84th percentiles of the SFR distribution in a given stellar mass bin) are marked by orange solid and dashed lines, respectively. G6025 is marked as a red star. The tight sequence of galaxies on the SFR-$M_{\rm{star}}$ plane is in part driven by the fact that both SFR and $M_{\rm star}$ are determined simultaneously from the fitting, and both are sensitive to the overall normalization of the SED. In the stellar mass bin of $\log{[M_{\rm{star}}/M_\odot]}=10.2-10.3$, the median SFR value in the field is $63^{+22}_{-15}~M_\odot$yr$^{-1}$, in good agreement with the prediction of a semi-analytic model by \citet{dutton10}. In comparison, the SFR estimate  for G6025  is $95\pm29~M_\odot$yr$^{-1}$. While the SFR of G6025 is $\approx $50\% higher than the median, but it is only marginally above the $1\sigma$ scatter on the upper end. Thus, despite being unusually large in its spatial extent, the SFR of G6025 is not unique compared to star-forming galaxies of similar stellar mass at high redshift.

\subsection{Size and Angular Structure of G6025}\label{angular_dist}

Typical high-redshift galaxies have  angular sizes that are only a small fraction of the PSF of imaging data taken in non-AO ground-based observations \citep[e.g.,][]{mosleh11,ribeiro16}.  G6025 is a clear exception to this limitation as it extends $\approx$3\arcsec\ across well beyond the size of their seeing disks. To characterize its angular structure, we use the original $Y$-band image, which has the best image quality with the measured stellar FWHM of 0.57\arcsec (4.2~kpc at $z=3.72$). In the left panel of Figure~\ref{fig_contour}, we show the contour lines corresponding to  the pixel intensities of 1.5$\sigma$, 2$\sigma$, 3$\sigma$, and 4$\sigma$. The pixel-wise rms noise $\sigma$ is computed on a 15\arcsec$\times$15\arcsec\ image subsection centered on G6025, with several iterations of sigma clipping.  The linear size of the 1.5$\sigma$ isophote along the major axis  is 3.2\arcsec\ or $24$~kpc. For reference, the size of the seeing disk is indicated as a grey circle in the figure.

In the surface brightness map, there appear to be several ``knots'' embedded within the 1.5$\sigma$ isophote, and they are aligned along the major axis. The two highest peaks ($\geq 4\sigma$) are $\approx$1.5\arcsec\ apart, but each is surrounded by more diffuse light on both sides. The separation of the two highest peaks is consistent with that in the 2D spectrum. 

We repeat the isophotal size measurement after smoothing the image with a gaussian kernel with a FWHM of 1.2 pixel (0.2\arcsec), which is sufficient to erase all  substructures seen in the original image. The sky rms is recomputed after the smoothing. The size is measured to be $\approx$3\arcsec\, consistent with that based on the original data (Figure~\ref{fig_contour}, middle). 

We use the GALFIT software \citep{galfit1,galfit2} to explore both single- and two-component scenarios. We use the $Y$-band sky-subtracted image centered on G6025. As initial guesses of source positions, we provide the image center (for the single-component fit) and two pixel coordinates located 0.5\arcsec\ from the image center (for the two-component fit). Our results are insensitive to the initial conditions of our GALFIT run.   The best-fit source positions and structural parameters are listed in Table~\ref{tab:galfit}; the latter includes S\'{e}rsic index $n$ ($S(r)\propto \exp{[r^{1/n}]}$), ellipticity $\epsilon$ ($\equiv 1-b/a$), and galaxy scalelength where $a$ and $b$ are semi-major and semi-minor axis, respectively. As for the galaxy size, we provide the effective radius $r_e$ (defined as $\sqrt{ab}$) for the general $n\neq 1$ case, and the disk scalelength $r_s$ for the exponential disk ($n=1$). Given the noisy data, it is not surprising that all models provide similarly good fits to the data with the reduced $\chi_r^2$ of 0.94. In all cases, the  galaxy is very elongated with the ellipticity of $\epsilon \geq 0.65$. The best-fit models and corresponding residual images are shown in Figure~\ref{fig_contour}.

\begin{table*}
\caption{Best-fit GALFIT parameters for G6025}
\begin{center}
\begin{tabular}{ccccc}
\hline
\hline
 & single & single & 2 component  & 2 component\\
 &   S\'{e}rsic index  &åExp disk    & S\'{e}rsic index & Exp disk \\ 
\hline
$r_e$ (half-light radius) &  1.19\arcsec (8.8~kpc) & 0.66\arcsec (4.9~kpc) & 0.44\arcsec (3.3~kpc)  & 0.30\arcsec (2.2~kpc)  \\
  &   & & 0.84\arcsec (6.2~kpc) & 0.45\arcsec (3.3~kpc) \\
$a$ (semi-major axis) & 2.89\arcsec (21.3~kpc)& 1.69\arcsec (12.5~kpc) & 0.75\arcsec (5.6~kpc) & 0.60\arcsec (4.4~kpc)\\
    &               &      & 1.67\arcsec (12.3~kpc) & 1.09\arcsec (8.1~kpc)\\
$n$ (S\'{e}rsic  index) &  0.13 & 1.00 (fixed) & 0.64 & 1.00 (fixed)\\
 &   & & 0.04 & 1.00 (fixed) \\
 $\epsilon$ (ellipticity) & 0.83 & 0.85& 0.66 & 0.75 \\
& - & -& 0.74 & 0.83 \\
Source Position$^\ast$ & (-0.00\arcsec,-0.06\arcsec)  & (-0.06\arcsec,-0.10\arcsec)  & (-0.70\arcsec,-0.69\arcsec) & (-0.67\arcsec,-0.65\arcsec) \\
 & -  & - & (0.47\arcsec,0.32\arcsec) & (0.47\arcsec,0.28\arcsec) \\
Separation & - &  - & 1.55\arcsec (11.5~kpc) & 1.47\arcsec (10.9~kpc)  \\
Flux ratio$^\dagger$ & - & - &   1.32 & 1.39  \\
\hline
$\chi_r^2$ &  0.938 & 0.942 & 0.937 & 0.937  \\
\hline
\end{tabular}
\end{center}
{$\ast$}~{Measured from image center. Image orientation is north up and east to left.\\}
{$\dagger$}~{Inferred flux ratio for two-component scenarios.}
\label{tab:galfit}
\end{table*}

The single-component model returns the largest effective radius:  when S\'{e}rsic  indices are left as a free parameter, we find the effective radius of  8.8~kpc and $n=0.13$. For an exponential disk model, we find 4.9~kpc.  The very large size found for the former is due to the fact that a low S\'{e}rsic index allows a large portion of the outer part of the galaxy to lie below the surface brightness limit of the image. Given that we cannot robustly constrain the S\'{e}rsic index from the data, we adopt the exponential disk profile as our fiducial model, which returns more conservative size estimates.  

Assuming two  exponential disks, the best-fit effective radii are 2.2 and 3.3~kpc, for the southeastern and northwestern source.  However, given the data, much larger galaxy sizes are equally likely.  When we fit the image with two sources with floating S\'{e}rsic indices, the best-fit solutions return larger sizes and smaller $n$. Thus, we are unable to determine how many sources G6025 is comprised of based on the current data. 

In the two-component case, the separation and the flux ratio are measured to be $\sim1.5$\arcsec\ (11~kpc) and 1.3-1.4, respectively. These quantities are well constrained regardless of the assumed light profile.  Given  the high ellipticities ($\equiv 1- b/a$: $0.7-0.8$) and measured semi-major axes of the galaxies (4.4~kpc and 8.1~kpc), the model fits suggest that the two sources are aligned in the same direction with substantial overlap. The inferred UV continuum luminosities are $(2.7-2.8)L^*_{\rm{UV}}$ and $(2.0-2.1)L^*_{\rm{UV}}$, where the more luminous source lies northwest of the other. We adopt the UV luminosity function measurements of \citet{bouwens07} at $z\sim4$.   Assuming that rest-frame UV and optical emission are cospatial, we estimate the stellar mass surface density; it is $\log{\Sigma_*} = 7.87^{+0.11}_{-0.14}$ ($M_\odot$~kpc$^{-2}$) for the single-source scenario, and $\log{\Sigma_*}  = 8.49^{+0.11}_{-0.14}$ and $8.70^{+0.11}_{-0.14}$ for the two-component exponential disk model. The errors only reflect the uncertainties in stellar mass estimate, and do not include those in size measurements. These estimates should be taken as a strict lower limit as all models considered here assume a smooth light and mass distribution. 

\begin{figure*}
\epsscale{1.0}
\plotone{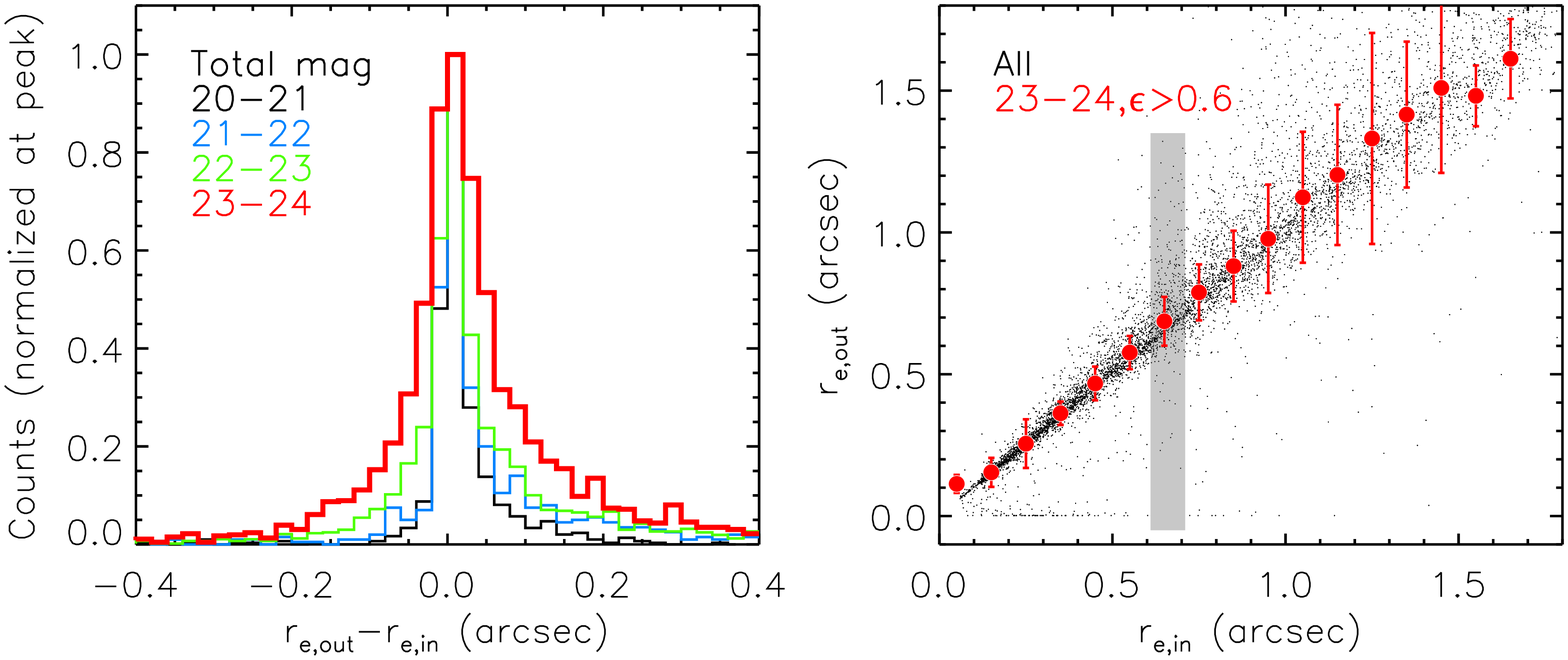}
\caption{
The robustness of galaxy size measurement is quantified through image simulations where artificial galaxies of known brightnesses and sizes are inserted into a mock image and the size measurements are carried out using GALFIT. 
{\it Left:} the deviation of the measured sizes relative to the true sizes is shown as histogram for the four magnitude groups, $20-21$ (black), $21-22$ (blue), $22-23$ (green), and $23-24$ (red). G6025 belongs to the last group. 
{\it Right:} input vs output galaxy sizes are shown for all galaxies (black) and for those with similar brightness and ellipticity to G6025 (red). The best-fit $r_e$ size for G6025 is marked as a grey region. The fractional error in size measurement for G6025 is $\sim$10\%.  
}
\label{fig_galfit}
\end{figure*}

When we run GALFIT on the $R$-band image, which is the deepest optical image but with poorer seeing than the $Y$ band, we obtain results in good agreement with those measured in the $Y$ band: the half-light radius of 1.26\arcsec\ for the S\'{e}rsic profile ($n=0.1$) and 0.68\arcsec\ for the exponential profile, respectively. However, our two-component model fits do not converge on the $R$ band. It is likely that the image is too blurred to provide a unique solution.

To evaluate the robustness of our size measurement, we perform image simulations. We create an empty image (500 pixels on a side) with noise properties identical to the $Y$-band image, and insert five galaxies at a time each with a known size and brightness. For all galaxies, we assume an exponential profile, $I\propto \exp{[-r/r_{e,{\rm in}}]}$ where the size $r_{e,{\rm in}}$ and magnitude range in $0.1\arcsec - 1.8$\arcsec\ and $20-24$~AB, respectively; the values for G6025 are 23.5~AB and $r_{e}=0.66$\arcsec.  In each of the 2,500 runs (simulating 12,500 galaxies in total), we use SExtractor-measured positions as the initial guess and run GALFIT to estimate the sizes.

 As illustrated in Figure~\ref{fig_galfit}, our ability to measure the galaxy sizes depends on the surface brightness, i.e., the combination of total flux, intrinsic size, and inclination angle.  For very large galaxies with low ellipticity (face-on), its low surface brightness causes the GALFIT-derived positions to often deviate significantly from the actual positions in which we insert the galaxies. Even when a large galaxy is recovered in the correct positions (which we define as within 1 pixel from the input positions), the scatter is significant and skewed towards larger sizes (right panel) as a part of the galaxy isophote can `hide' underneath the noise.  Given everything equal, high ellipticities help increase the surface brightness level, making it easier to recover the intrinsic size.  Our simulation indicates that our size measurement should be relatively robust. When galaxies of similar brightnesses ($23-24$~AB), sizes ($\sim 0.7$\arcsec), and ellipticities ($\epsilon \geq 0.5$) are considered, we find the deviation $\Delta r_e$ to be $0.02\pm0.06$\arcsec\ where $\Delta r_e \equiv r_{e, {\rm out}} - r_{e,{\rm in}}$. Thus, the fractional error in size measurement is roughly 10\%. 

\begin{figure*}[ht!]
\epsscale{1.0}
\plotone{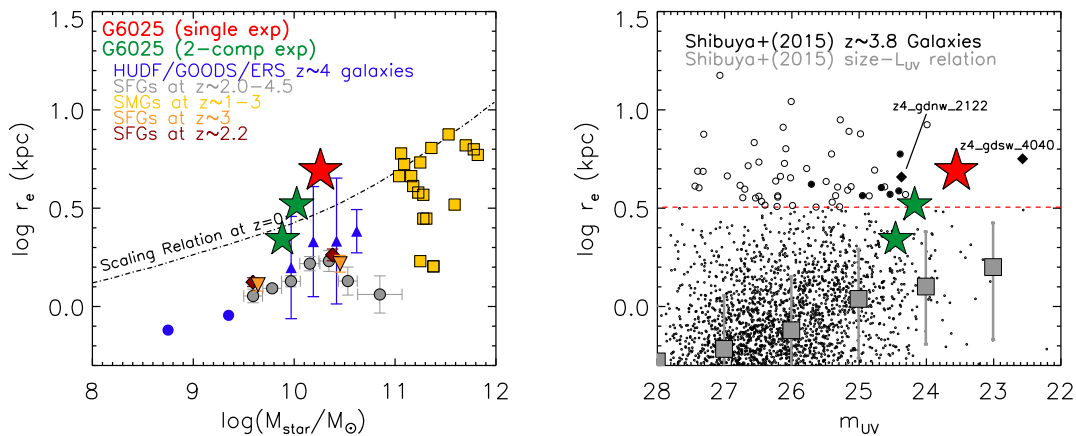}
\caption{
Compilation of galaxy size measurements in the literature. All measurements are based on HST images with the exception of G6025.  {\it Left:} the stellar mass-size scaling relation. Yellow squares show the submillimeter galaxies studied by \citet{targett13} while other symbols show those of normal star-forming galaxies at similar cosmic epoch reported by \citet[][blue triangles]{mosleh11}, \citet[][blue circles]{mosleh12}, \citet[][grey circles]{ribeiro16}, and \citet[][orange triangles and brown diamonds for $z\sim2.2$ and $z\sim3$, respectively]{law12}. 
The  relation at $z\sim0$ are shown as dashed-dot line \citep{shen03}. The size obtained assuming a single component (red star) exceeds the local scaling law. The two-component exponential disk model (labelled `2-comp exp') yields smaller galaxy sizes that are in better agreement with the observed scaling law. However, the model still requires the brighter galaxy to be $\sim$40\% larger than the expected median. 
{\it Right:} The size measurements of a $z\sim4$ galaxy sample compiled by \citet{shibuya15} are shown as small circles; the size-$L_{\rm UV}$  scaling relation given by \citet{shibuya15} is denoted by large grey squares. A small subset (0.7\%) of galaxies have the sizes $r_e\geq4$~kpc similar to G6025 (orange circles). 
}
\label{fig_sizes}
\end{figure*}

In Figure~\ref{fig_sizes}, we show the compilation of  size measurements of galaxies at similar redshift \citep{mosleh11, mosleh12, law12, shibuya15, ribeiro16}. All measurements are derived from  GALFIT fits based on various HST datasets, even though their depths and wavelengths vary. Only G6025's size is measured from the ground-based image.  Some of these measures are based on the HST/WFC3 data while others use the HST/ACS data, which sample the rest-frame near-UV/optical and far-UV wavelengths, respectively. However,  \citet{shibuya15} noted that the size measurements based on the ACS and WFC3 data are essentially identical, i.e., no significant morphological $k$-correction is required at high redshift \citep[also see][]{law12}. 

The galaxies whose sizes are most similar to that of G6025 are the submillimeter galaxies  (SMGs) studied by \citet{targett13}, which are heavily dust-obscured starburst galaxies. The median (mean) size of the nineteen galaxies at $z=1.1-2.8$ ($\langle z \rangle=2.0$) is 4.42 (4.60)~kpc\footnote{The values quoted here represent the S\'{e}rsic parametric fit to all components. However, as noted by  \citet{targett13}, the size measurements done on the brightest component are only slightly smaller.} with the standard deviation of 1.8~kpc. The largest (smallest) galaxy in their sample has the half-light radius of 7.5 (1.6)~kpc.  In comparison, most UV-luminous star-forming galaxies with moderate dust are characterized by uniformly smaller sizes than that of G6025 by at least a factor of 3. 

Despite their similar sizes, SMGs and G6025 are not similar systems. 
\citet{targett13} noted that despite frequent occurrences of multiplicities in their sample galaxies, there is usually one component that dominates the total flux. Even if G6025 is composed of two components, their brightness levels would be comparable. We find that the inferred separation between the two brightest knots within G6025 ($\approx 1.5$\arcsec) is not unusual when compared with that of SMGs with multiplicities. Finally, all of the SMGs are much more massive systems with their median stellar mass  nearly an order of magnitude higher than G6025. 

As illustrated in the left panel of Figure~\ref{fig_sizes}, there is a clear trend that galaxy sizes increase with increasing stellar mass \citep{franx08}. 
When compared to the same trend at $z\sim0$ \citep{shen03}, the redshift evolution of the relation is also apparent. Despite their large sizes, SMGs lie typically on or slightly above the $M_{\rm{star}}$--$r_{\rm e}$ relation extrapolated from other measurements\footnote{While \citet{targett13} measured the sizes of individual components, they also give the `total' size by fitting all multiple component sources with a single-component model. Since their stellar masses are estimated from ground-based data in which individual clumps are not resolved, we show their total sizes in this figure.}. Given that these SMGs generally lie at lower redshift than blue star-forming galaxies (of the nineteen galaxies considered here, 18 (95\%) lie at $z\leq 2.5$ and 10 (53\%) lie at $z \leq 2.0$), and galaxy sizes increase with cosmic time, it is possible that SMGs obey the same power-law scaling relation as their bluer star-forming cousins.

The single-component size of G6025 exceeds not only the expectation from the scaling relation at high redshift  but also that of local galaxies. Given its stellar mass, it is larger by a factor of $\gtrsim3$ than the expectation at high redshift. We also consider the sizes obtained from our two-component fit. We assume the uniform stellar-mass-to-UV-light ratio and use the best-fit flux ratio to estimate the stellar mass of each disk. The smaller galaxy sizes inferred in this scenario bring both components to a much better agreement with the scaling relation than previously. They lie at the upper end of the 1$\sigma$ range of star-forming galaxies whose UV luminosities are comparable to that of G6025 \citep{mosleh11}.

 A similar conclusion can be drawn when the $m_{\rm UV}$--$r_e$ scaling relation is considered. In the right panel of Figure~\ref{fig_sizes}, we show the size measurements of a large sample of LBGs at $z\sim4$ compiled by \citet{shibuya15} as a function of apparent magnitudes in the bands sampling the rest-frame wavelength $\lambda_{\rm rest}\approx 1700$--1800\AA.  Large squares denote the mean scaling relation given by \citet{shibuya15} after being adjusted to reflect the adopted cosmology. Once again, G6025 lies well above the scaling relation. 
 
Of the 2,890 galaxies in the \citet{shibuya15} size catalog, we find 55  galaxies larger than $r_e= 3$~kpc. The two brightest  galaxies are {\tt z4\_gdsw\_4040} and {\tt z4\_gdnw\_2122}, which have the total magnitudes in the $z_{850}$ band of 22.57 and 24.37, respectively; we identify their sky positions given in the \citet{harikane16} catalog and use the GOODS v2.0 cutout service\footnote{{\tt https://archive.stsci.edu/eidol\_v2.php}} to inspect their morphologies. The  source {\tt z4\_gdsw\_4040} ($\alpha$= 53.1935, $\delta$=27.8988 (J2000); $r_e$=5.6~kpc, $q=0.34$) has a linear morphology with a uniform surface brightness across its face, perhaps is most similar to G6025. The source {\tt z4\_gdnw\_2122} is a long chain galaxy ($\alpha$=189.2107, $\delta$=62.1397 (J2000); $r_e$=4.6~kpc, $q$=0.37); its light is dominated by several point sources surrounded by diffuse emission. We also inspect the remaining 53 galaxies. There are several galaxies that are very extended but at the low surface brightness level. These galaxies are indicated by black filled circles in Figure~\ref{fig_sizes} (right). However, we find that for the many of the sources (open circles in Figure~\ref{fig_sizes}), their large sizes are likely a result of a poor fit due to the proximity to a much brighter source or a diffraction spike, or due to source crowding.  

Our analysis suggests that large and luminous galaxies at high redshift are exceedingly rare, but do exist. Without the higher resolution morphological data, it remains unknown which class of the large galaxies G6025 is closest to: SMGs, chain galaxies, or a uniformly bright and long galaxy such as {\tt z4\_gdnw\_2122}. Another interesting question is whether or not these large galaxies share a common origin  observed at a different stage, or multiple physical processes are responsible for their formation.  A uniform analysis of their resolved kinematics and stellar population studies will be necessary to further elucidate their nature. 


\subsection{Location and Environment}
\begin{figure*}
\epsscale{1.1}
\plotone{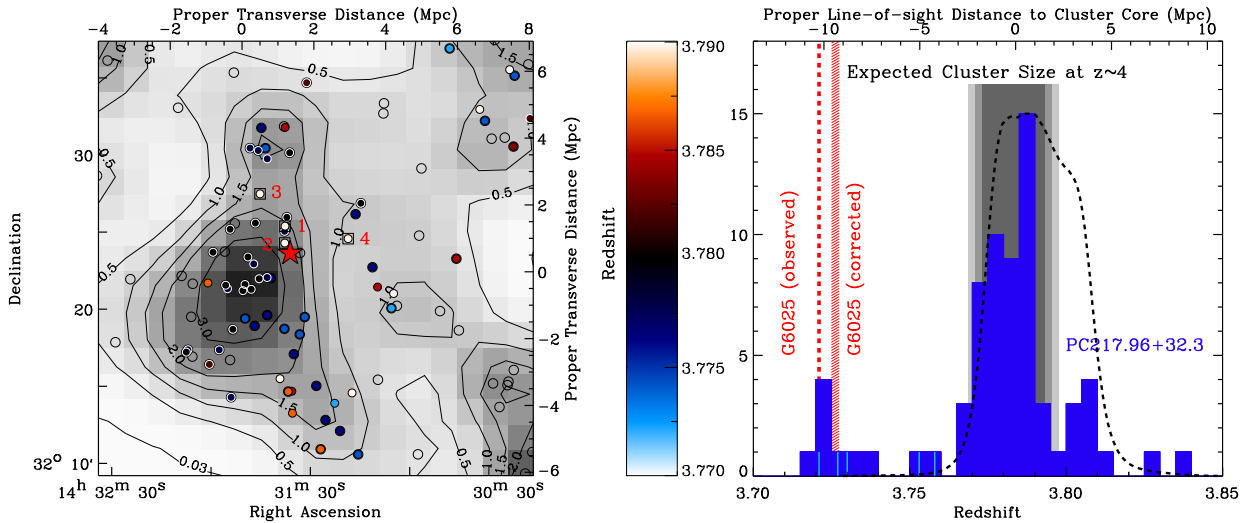}
\caption{
{\it Left: } The position of G6025 (red star) and of the galaxies in and around the protocluster PC217.96+32.3. Spectroscopically confirmed sources are shown as filled circles color-coded by redshift according to the color bar.
All galaxies in the redshift range $z=3.717-3.761$ are shown in white; four of these sources closest to G6025, which are presumably companion galaxies, are  marked with open squares with their local IDs in red.  Photometric LAEs are shown in open circles. Contours and shades represent the relative enhancement of LAE surface density constructed by smoothing the positions of all LAEs by a 2D gaussian kernel (FWHM=4.7\arcmin\ or 10~Mpc comoving). {\it Right:} the redshift histogram in the  PC217.96+32.3 field (blue). Red dashed line marks the  G6025 redshift and the corresponding proper line-of-sight distance from the protocluster. A hatched red region shows the  separation after  correcting for the redshift offset (see text). The redshifts of G6025 and its four companions are marked in cyan lines. Three grey shades mark the expected extent (median and $\pm 1\sigma$) of a Coma progenitor  at $z\sim4$  \citep{chiang13}. The redshift selection function for photometrically selected LAEs (dashed line) does not extend to the redshift of G6025 and its companions.
}
\label{fig_location}
\end{figure*}

G6025 is located in the sightline of the PC217.96+32.3 protocluster \citep{lee14,dey16}. In the left panel of Figure~\ref{fig_location}, we show the angular distribution of all known galaxies at the protocluster redshift ($z=3.785$) together with the position of G6025. The relative enhancement of the source density is estimated using the angular positions of LAEs only (including photometric LAE candidates) and is shown as contour lines and shades.  

In the right panel of Figure~\ref{fig_location}, we show a histogram of spectroscopic redshifts in our survey field. The median (mean) redshift of all protocluster member galaxies is $z=3.783$ (3.781). Adopting the median (mean) redshift as the protocluster redshift, the line-of-sight distance between G6025 and PC217.96+32.3 is 10.2 (9.9)~Mpc. However, the actual proper distance is slightly smaller given the fact that interstellar absorption and Ly$\alpha$ emission line centroids are typically blueshifted and redshifted relative to stars, respectively. 
By assuming that G6025 and protocluster LAEs are typical galaxies of their respective types, we  correct their redshifts accordingly to estimate their systemic redshift, $z_{\rm{sys}}$. For LAEs, we use the measurements of \citet{erb14} made on two statistical samples of LAEs at $z\sim2.3$ and $z\sim3.1$; at the continuum luminosity comparable to galaxies in our sample, the Ly$\alpha$ offset is roughly  $\Delta v_{\rm{Ly}\alpha}\sim150$~km/s (or $z_{\rm{sys}} = z_{\rm{Ly}\alpha} - 0.0024$). Similarly, we use the calibration given in \citet{steidel10} to correct for the G6025 redshift measured from the interstellar lines by $ z_{\rm{sys}}=z_{\rm{IS}}+0.00278$. The corrected line-of-sight distance from G6025 to protocluster is then 9.3 (9.0)~Mpc (physical). 

In the transverse direction, G6025 lies at the outskirts of the extended LAE overdensity. The core, defined as the center of 10 galaxies at the peak LAE surface density, is 3.1\arcmin\ away from G6025, corresponding to 1.4~Mpc (physical). The three-dimensional proper distance from G6025 to the protocluster center is 9.1-9.4~Mpc. 

Given its distance from the core of PC217.96+32.3, it is unlikely that G6025 is part of the protocluster. \citet{chiang13} studied the spatial distribution of all galaxies -- well before the coalescence -- that would end up as members of a present-day galaxy cluster; they determined that, when observed at $z\sim 3-4$, the progenitors of the most massive clusters ($M_{\rm{total}}\gtrsim 10^{15}M_\odot$) would be characterized with the effective radius in the range of $2.0\pm0.4$~Mpc. The angular extent of PC217.96+32.3  defined by the spectroscopic redshift spike already fits comfortably in this range  (grey shades in Figure~\ref{fig_location}, right). Several other protocluster structures have similar sizes \citep[e.g.,][]{matsuda05}. The density enhancement at $\sim 9$~Mpc away from the center of the parent halo cannot be more than a few percent \citep[][]{chiang13}.

\begin{table*}
\caption{Locations and key characteristics of G6025 and its closest neighbors}
\begin{center}
\begin{tabular}{lccccccccc}
\hline
\hline
ID & ID$_{\rm spec}$ & $\alpha_{\rm J2000}$ & $\delta_{\rm J2000}$ & $z_{\rm spec}$ & Distance  & $I_{\rm AB}$ & $M_{\rm UV}$ & $B_W-R$ & $R-I$ \\
\hline
\#0 & G6025 & 217.9000 & 32.3940 & 3.721 & - & $23.55$ & $-22.42$ & $>3.2$ & $0.18\pm0.14$ \\
\hline
\#1 & BD46750 &  217.9066 & 32.4231 & 3.727 & $<$1.3~Mpc  & 25.23 &  $-20.74$  & $2.79\pm0.58$ & $0.05\pm0.18$ \\
\#2 & NDWFS\_Bw\_99073 & 217.9069 & 32.4049 &  3.730 & 1.6~Mpc & 24.16 & $-21.82$ & $2.61\pm0.29$ & $0.29\pm 0.08$  \\
\#3 & NDWFS\_Bw\_10277 &  217.9380 & 32.4581 & 3.753 & 5.7~Mpc  & 24.10 & $-21.89$  & $3.81\pm 0.10$ & $0.40\pm0.10$\\
\#4 & BD47884 &  217.8269 & 32.4095 & 3.758 & $<$6.5~Mpc  & 25.00 & $-20.99$ & $2.94\pm 0.98$ & $0.48\pm0.20$\\
\hline
\end{tabular}
\end{center}
\label{tab:neighbor}
\end{table*}

To examine the local environment of G6025, we search our current spectroscopic catalog for galaxies within the redshift range  of $z=3.721\pm 0.04$, which covers a 13~Mpc line-of-sight distance from $z=3.721$. We find 11 LBGs, 9 of which are within the image shown in Figure~\ref{fig_location} (left) where their locations are indicated by white circles. Of those, we mark the four LBGs which may be associated with G6025 as boxes and label them with numbers 1--4 ordered by  redshift. Galaxy \#2 and \#3 have very luminous UV continua ($\approx (1.7-1.8) L^*_{\rm{UV},z\sim4}$) from which we measured interstellar absorption redshifts. On the other hand, galaxy \#1 and \#4 are sub-luminous galaxies ($(0.6-0.8) L^*_{\rm{UV},z\sim4}$) whose redshifts were identified through weak Ly$\alpha$ emission. In Table~\ref{tab:neighbor}, we  list their  spectroscopic redshifts, coordinates, UV brightness, and approximate distances to G6025. The distance computed using the Ly$\alpha$ redshift sets an upper limit, and is indicated as such in the table.

While five galaxies (counting G6025 as one galaxy) may not seem impressive compared to the redshift overdensity of PC217.96+32.3, it presents a strong evidence that G6025 lives in a dense environment, given the following considerations. Much of the existing spectroscopy in the field was aimed at redshift confirmation of photometrically pre-selected LAEs via Ly$\alpha$ emission \citep[][]{dey16}, and  as a result, lack sufficient depths to identify sources that are not strong Ly$\alpha$ emitters. Only a subset of relatively bright LBGs \citep[see][for details]{lee13} near the LAE overdensity were observed as secondary targets, and the majority that yielded redshift measurements were due to the presence of Ly$\alpha$ emission except for a few that were exceptionally bright. Additionally, it is worth noting that a majority of known members of PC217.96+32.3 are LAEs; a large fraction of them are not formally detected in the $I$ band which samples the rest-frame UV continuum near 1700\AA. We have no way of identifying similarly UV-faint galaxies near G6025 as there is currently no suitable narrow-band filter to do so.  Without a systematic survey of UV-luminous galaxies in the entire field, it is difficult to quantify the significance of having found four galaxies near G6025. After excluding all galaxies in the protocluster redshift, we search for similar LBG groups and find none. Thus, we argue with reasonable confidence that G6025 inhabits a locally overdense environment. 
 
\section{What is the physical nature of G6025?}
With the limited data currently available, we can only speculate about the physical origin of G6025 and its dynamical state.  
One possibility is that G6025 is a high-redshift analog of of massive and giant disk galaxies found in local universe \citep{ogle16}. The fact that both spectral and imaging data show the presence of two dominant off-center components strongly suggests that G6025 is not a large classic disk viewed edge-on. However, it is possible that the observed rest-frame UV morphology may be caused by a very compact dusty core  rather than the presence of two stellar clumps. This can be unambiguously determined from the rest-frame infrared morphology -- past the rest-frame 1.6$\mu$m stellar bump ($\lambda_{\rm obs}\gtrsim 8$~$\mu$m) -- where the stellar emission is virtually free of dust obscuration.

Alternatively, G6025 may represent a clumpy thick disk in formation. Individual kiloparsec-scale clumps are presumably formed through gravitational instability in  gas-rich turbulent disks \citep[e.g.,][]{dekel09}. Depending on the viewing angle, the clumps can appear strung together in a line as a `chain' galaxy, or more distributed in the angular space as a `clump cluster'. The nearly linear structure of G6025 and multiple unresolved intensity peaks  in the surface brightness map (Figure~\ref{fig_contour}) are qualitatively similar to those observed for  chain galaxies. 

Samples of clumpy galaxies were identified in deep HST surveys such as the Hubble Ultradeep Field \citep[HUDF:][]{hudf} and the Cosmic Assembly Near-infrared Deep Extragalactic Survey \citep[CANDELS:][]{grogin11,koekemoer11}, and their physical properties have been characterized in a  statistical manner \citep[e.g.,][]{elmegreen05clump,elmegreen07,wuyts12, guo12, guo15, guo17}.
There is no general consensus in the literature on the fate of these clumps: e.g., they could migrate to the center and form a young stellar bulge of a disk galaxy \citep[e.g.,][]{bournaud07,dekel09,bournaud14}, or dissipate into  the disk due to feedback and tidal interactions \citep{murray10,genel12,buck17}. 
However, existing studies agree that the clumps have higher specific SFRs and are generally younger than the more diffuse `disk' component \citep{guo12, wuyts12}.  Individual clumps typically have sizes $\lesssim$1~kpc, which would not be resolved in our ground-based data.    

If G6025 is a clumpy chain galaxy, it would be the largest one known. We compare its size with the sample studied by \citet{elmegreen05}. They classified  $\sim$120 as chain galaxies in the HUDF, of which about 30 galaxies lie at $z_{\rm{phot}}=3-5$ \citep{elmegreen07}. The sizes of these galaxies are defined as half the separation of the extent of its linear structure, a metric size more appropriate for clumpy structures with no clear center than using S\'{e}rsic profiles (surface brightness $\propto \exp{(-r^{1/n})}$). \citet[][see their Figure~12]{elmegreen07} show that the median size of  20 chain galaxies  and that of 17 tadpole galaxies at $z_{\rm{phot}}$=3-5 are comparable at $\sim2$~kpc. Adopting a similar size measurement on G6025, correcting for the PSF effect to the best of our ability, gives the size of $\approx$10~kpc, a factor of 5 larger than the median of the HUDF galaxies. The largest chain and tadpole galaxy in the Elmegreen et al. sample are 4.5~kpc and 6.0~kpc in size, respectively, i.e., roughly half the size of G6025. 

We also compare the sizes measured from a larger clump sample identified by \citet{guo17}, which contains 132 $z \geq 2$ galaxies with two or more clumps. They do not use the Elmegreen et al. size metric, but instead adopt GALFIT-measured semi major axis sizes; the mean (median) value for the clumpy galaxies is 3.4 (3.1) kpc with a standard deviation of 1.3~kpc. All the Guo et al. sample galaxies are $<$11~kpc in size with the three largest galaxies in the range of 7--11~kpc.  In comparison, G6025 has a semi-major axis size of 21.3 (12.5)~kpc estimated using the GALFIT single-S\'{e}rsic (exponential profile: $n=1$) fit to the light profile. 

G6025 is also much more luminous than a typical chain galaxy. The median  $z_{850}$-band magnitude of chain or tadpole galaxies is 27~AB. Interpolating between the $I$ and $Y$-band photometry, the apparent magnitude of G6025 in the same band is expected to be 23.5~AB, a factor of 25 more luminous than known chain galaxies. There is little uncertainty in our interpolation as $I-Y\sim0.03$. The brightest chain galaxy and tadpole galaxy in their sample have $\sim$25 and $\sim$24~AB, respectively, a factor of 2-4 times fainter than G6025. 

 A small subset of clump cluster galaxies match the size and apparent brightness of G6025 most closely. \citet{elmegreen05clump} selected the ten largest and brightest clump cluster galaxies in the HUDF, and found the average diameter of 19~kpc  and the $i$ band brightness in the range of $22.0-24.5$. Of those, we identify two sources that are possibly in a similar configuration to G6025, i.e., a linear structure showing two dominant clumps separated by 1\arcsec--2\arcsec; their IDs given in \citet{elmegreen05clump}  are 3034+ and 3465+ (see their Figure~1). The remaining eight have axial ratios closer to unity and are generally dominated by a larger number of clumps. We cross correlate their sky positions with the MUSE HUDF spectroscopic catalog \citep{inami17}, and find seven matches. The spectroscopic redshifts for the objects 3034+ and 3465+ are 2.678 and 1.766, respectively. The other five clump cluster galaxies with redshift identification lie at $z=1.43-1.77$. While they are genuinely very large galaxies, most of them are observed when the universe was  $\gtrsim2$~Gyr older than when  G6025 is observed. 
Furthermore, considering the expected cosmological dimming (1.4 -- 2.0~mag assuming a flat SED, $f_\nu\propto \nu^0$) and k-correction, they are not as luminous as G6025.

Finally, G6025 may be comprised of two interacting galaxies caught in the early stage of their merger. In this scenario, the separation between the two sources ($\approx$11~kpc; Table~\ref{tab:galfit}) suggests that they are nearly aligned along their semi-major axis with significant overlap. Thus, the observed light profile may be a combination of two disks and tidal features.  A slight inflection at the southeastern end of the galaxy (seen as a $3\sigma$ intensity peak in Figure~\ref{fig_contour}) suggests that the structure may not be entirely linear. The ubiquity of tadpole galaxies, doubles, and other `trainwreck' galaxies at high redshift shows that  mergers are common ways for galaxies to assemble their masses at this epoch \citep[e.g.,][]{lotz11}. The linear morphology with two cores as observed in G6025 is qualitatively similar to that seen in hydrodynamical simulations of an equal-mass merger of two gas-rich disk galaxies near the first-pass closest approach \citep[see Figure~2 of][]{lotz08}. The fact that G6025 lies only slightly above the main sequence  does not preclude the possibility of merger. \citet{lotz08} found that, while  the morphology of the merging  system is very clearly disturbed, the enhancement of star formation acitivities is a much more prolonged process, only to peak 1~Gyr later. \citet{kaviraj13} argued that the SFR enhancement due to major merger is at best modest \citep[also see,][]{fensch17}.

The scenarios considered here can be tested with new observations. Deep, spatially resolved spectroscopy will be able to definitively test the existence of a very large disk. The brightest emission line, H$\alpha$, redshifts to 3.1$\mu$m, which will require future JWST/NIRSPEC observations; however, given the low reddening, it might be possible to use [O\,{\sc ii}]$\lambda\lambda$3727 and [O {\sc iii}]$\lambda\lambda$4959,5007, which move to the edge of the $H$ and $K$ band, respectively, to trace the kinematics.  The measured velocity dispersion and the spatial velocity profile would provide a firm metric to characterize the similarity of its kinematics  to normal-sized star-forming galaxies at high and low redshift. 

Detailed morphologies from higher angular resolution data will be able to discern internal stellar structures by  detecting stellar clumps and/or tidal features, if present, and determine whether the galaxy is a giant edge-on disk. The number of clumps would provide a useful constraint. \citet{guo12}  found the median number of clumps per galaxy of {\it four}, in their study of clumpy galaxies in the HUDF. Finding three or more clumps would thus strongly argue against the alternative scenario of galaxy merger as three-way merger should be extremely rare. The caveat, however, is that a lower number of clumps does not necessarily favor the merger scenario. Based on a larger sample of clumps (identified from a set of shallower data than the HUDF),  \citet{guo17} found the median (mean) number of clumps of 2.0 (2.6) \citep[also see][]{ribeiro17}; the number would be even lower if systems with a single clump residing within more diffuse stellar light are also considered.  

Third, the galaxy's internal color gradient will enable more quantitative comparison with clumpy galaxies.   Recent studies of clumpy galaxies find that clumps tend to be younger, more massive, and less dusty than the inter-clump regions, while the central clumps ($d_{\rm{clump}} \lesssim 0.5 a$)  are older  and more massive than those in the outskirts \citep[e.g.,][]{guo17}. These trends are in a broad agreement with the expectation of the clump migration scenario. In the merger scenario  the physical properties should not have any strong radial dependence. Furthermore, the core of the two galaxies, which have evolved entirely independently,  are likely characterized by a different set of properties. However, their overall morphologies and color structures undoubtedly will change over the course of their interaction, and it is difficult to develop a clear set of expectation here. 

Finally, with the upcoming JWST/NIRCam, it will be possible to robustly test the presence of an extremely dusty core by sampling beyond the 1.6$\mu$m stellar bump feature (7.5 $\mu$m in the observer's frame). Regardless of its nature, G6025 provides a uniquely bright target to study one of the main modes of galaxy assembly at high redshift. 

\acknowledgments
We thank the anonymous referee for a careful reading of the manuscript and a thoughtful report, and Kristin Chiboucas at the Gemini Observatory for assisting GMOS observations.  KSL thanks Yicheng Guo for kindly sharing his clump catalog,  Masami Ouchi and Takatoshi Shibuya for providing their latest size measurements of high-redshift galaxies, and Hwihyun Kim for useful discussions in the early stage of this work. 
Based on observations at Kitt Peak National Observatory, National Optical Astronomy Observatory (NOAO Prop. IDs 2012A-0454, 2014A-0164, 2015A-0168, 2016A-0185; PI: K.-S. Lee), which is operated by the Association of Universities for Research in Astronomy (AURA) under cooperative agreement with the National Science Foundation. The authors are honored to be permitted to conduct astronomical research on Iolkam Du'ag (Kitt Peak), a mountain with particular significance to the Tohono O'odham. 
Data presented herein were obtained at the W. M. Keck Observatory using telescope time allocated to the National Aeronautics and Space Administration through the agency's scientific partnership with the California Institute of Technology and the University of California. The Observatory was made possible by the generous financial support of the W. M. Keck Foundation. The authors wish to recognize and acknowledge the very significant cultural role and reverence that the summit of Mauna Kea has always had within the indigenous Hawaiian community. We are most privileged to be able to conduct observations from this mountain. This work was supported by a NASA Keck PI Data Award, administered by the NASA Exoplanet Science Institute. We thank NASA for support, through grants NASA/JPL\# 1497290 and 1520350. 
This paper is based in part on data collected at Subaru Telescope, which is operated by the National Astronomical Observatory of Japan. 
This work is based in part on observations made with the Spitzer Space Telescope, which is operated by the Jet Propulsion Laboratory, California Institute of Technology under a contract with NASA.
AD's research was supported in part by the National Optical Astronomy Observatory (NOAO), which is operated by the Association of Universities for Research in Astronomy (AURA), Inc. under a cooperative agreement with the National Science Foundation.


\end{document}